\documentclass{aa}
\usepackage{graphicx}
\usepackage{txfonts}
\newcommand{\swift}  {Swift\,J0243.6+6124}
\newcommand{\ha}  {H$\alpha$}
\newcommand{\ew}  {EW(H$\alpha$)}
\newcommand{\ergs}  {erg s$^{-1}$}
\newcommand{\kms}  {km s$^{-1}$}

\def\simless{\mathbin{\lower 3pt\hbox
     {$\rlap{\raise 5pt\hbox{$\char'074$}}\mathchar"7218$}}}   
\def\simmore{\mathbin{\lower 3pt\hbox
     {$\rlap{\raise 5pt\hbox{$\char'076$}}\mathchar"7218$}}}   

\def\msun{~{\rm M}_\odot}
\def\rsun{~{\rm R}_\odot}
\usepackage{amstext}

\begin{document}

   \title{Optical counterpart to Swift\,J0243.6+6124}

   \subtitle{}
  \author{P. Reig \inst{1,2}
        \and
        J. Fabregat\inst{3}
        \and
        J. Alfonso-Garz\'on\inst{4}
           }

\authorrunning{}
\titlerunning{\swift}

   \offprints{pau@physics.uoc.gr}

   \institute{Institute of Astrophysics, Foundation for Research and
   Technology-Hellas, 71110 Heraklion, Greece 
         \and Physics Department, University of Crete, 71003 Heraklion, Greece
                \email{pau@physics.uoc.gr}
        \and Observatorio Astron\'omico, Universidad de Valencia, Catedr\'atico
        Jos\'e Beltr\'an 2, 46980 Paterna, Spain
         \and Centro de Astrobiolog\'{\i}a-Departamento de Astrof\'{\i}sica (CSIC-INTA), Camino Bajo del Castillo s/n,
        28692 Villanueva de la Ca\~nada, Spain
        }

   \date{Received ; accepted}

\abstract
{Swift\,J0243.6+6124 is a unique system. It is the first and only ultra-luminous
X-ray source in our Galaxy. It is the first and only high-mass  Be X-ray pulsar
showing radio jet emission. It was discovered during a giant X-ray outburst in
October 2017. While there are numerous studies in the X-ray band,
very little is known about the optical counterpart. } 
{Our aim is to characterize the variability timescales in
the optical and infrared bands in order to understand the nature of this intriguing system.  } 
{We performed optical spectroscopic observations to determine the spectral type.
Long-term photometric light curves together with the equivalent
width of the H$\alpha$ line were used to monitor the state of the circumstellar
disk. We used $BVRI$ photometry to estimate the interstellar absorption and 
distance to the source. Continuous photometric monitoring in the $B$ and $V$ bands 
allowed us to search for intra-night variability.
 }
{The optical counterpart to Swift\,J0243.6+6124 is a V = 12.9, O9.5Ve star, located at a distance of $\sim5$ kpc. The optical extinction in the direction
of the source is $A_V=3.6$ mag. The rotational velocity of the O-type
star is 210 km s$^{-1}$. The long-term optical variability agrees with the 
growth and subsequent dissipation of the Be circumstellar disk after the 
giant X-ray outburst. The optical and X-ray luminosity are strongly correlated
during the outburst, suggesting a common origin.
We did not detect short-term periodic variability that could be associated with nonradial pulsations from the Be star photosphere}
{The long-term optical and infrared pattern of variability of Swift\,J0243.6+6124 
is typical of Be/X-ray binaries. However, the absence of nonradial pulsations 
is unusual and adds another peculiar trait to this unique source. }

\keywords{stars: individual: \swift,
 -- X-rays: binaries -- stars: neutron -- stars: binaries close --stars: 
 emission line, Be
               }

   \maketitle

\begin{table*}
\caption{Log of the spectroscopic observations.}
\label{speclog}
\begin{center}
\begin{tabular}{lcc|cc|cc}
\noalign{\smallskip}    \hline \noalign{\smallskip}
Date    &JD (2,400,000+)  &Telescope    &Wavelength     &EW(H$\alpha$)&Wavelength   &EW(H$\beta$)  \\
        &                 &             &range ($\AA$)  &($\AA$)    &range ($\AA$)  &($\AA$)\\
\noalign{\smallskip}\hline\noalign{\smallskip}
04-10-2017  &58031.42   &SKO    &5460--6740     &$-11.0\pm0.3$  &3880--5315     &$-1.2\pm0.2$  \\
13-10-2017  &58040.37   &SKO    &5460--7380     &$-10.2\pm0.3$  &3690--5580     &$-1.3\pm0.2$  \\
15-10-2017  &58042.36   &SKO    &5460--7380     &$-9.8\pm0.3$   &3690--5580     &$-1.1\pm0.2$  \\
30-10-2017  &58057.49   &WHT    &6450--7260     &$-11.2\pm0.6$  &3830--4695     &--     \\
22-11-2017  &58080.26   &SKO    &5450--7370     &$-11.1\pm0.3$  &--             &--     \\
15-07-2018  &58315.58   &SKO    &5465--7375     &$-10.8\pm0.4$  &--             &--     \\
22-08-2018  &58353.56   &SKO    &5400--7300     &$-10.1\pm0.3$  &--             &--     \\
25-08-2018  &58356.54   &SKO    &5400--7300     &$-10.3\pm0.3$  &--             &--     \\
19-09-2018  &58381.42   &SKO    &5400--7300     &$-9.4\pm0.2$   &--             &--     \\
08-10-2018  &58400.51   &SKO    &5400--7300     &$-9.1\pm0.2$   &--             &--     \\
19-08-2019  &58715.59   &SKO    &5375--7285     &$-4.9\pm0.2$   &--             &--     \\
09-09-2019  &58736.53   &SKO    &5375--7285     &$-5.8\pm0.3$   &--             &--     \\
09-02-2020  &58889.40   &WHT    &6355--7255     &$-7.5\pm0.4$   &3850--4675     &--     \\
\noalign{\smallskip}    \hline
\end{tabular}
\end{center}
\end{table*}

\begin{table*}
\caption{Photometric observations of the optical counterpart to \swift\ from the Skinakas observatory.}
\label{photlog}
\begin{center}
\begin{tabular}{cccccc}
\noalign{\smallskip}    \hline\noalign{\smallskip}
\multicolumn{6}{c}{Photometry (mag.)}\\
\noalign{\smallskip}    \hline\noalign{\smallskip}
Date &  JD (2,400,000+)    &   $B$  &   $V$   &   $R$  & $I$   \\
\noalign{\smallskip}    \hline\noalign{\smallskip}
30-07-2019      &58695.584      &$13.86\pm0.01$ &$12.91\pm0.01$ &$12.24\pm0.01$ &$11.55\pm0.02$   \\        
11-09-2019      &58738.594      &$13.83\pm0.03$ &$12.86\pm0.01$ &$12.18\pm0.01$ &$11.45\pm0.02$   \\        
\noalign{\smallskip}    \hline\noalign{\smallskip}
\end{tabular}
\end{center}
\end{table*}

\section{Introduction}

\swift\ was first detected by the Burst Alert Telescope (BAT) on board the Neil
Gehrels {\it Swift} observatory on 3 October 2017  during a giant X-ray
outburst \citep{kennea17}. The timing analysis revealed a  periodicity at
9.86s, suggesting that the new transient is an X-ray pulsar. Pulsations were
subsequently confirmed by {\it Fermi} \citep{jenke17} and {\it NuSTAR} \citep{bahramian17}.
The Fermi/GBM accreting pulsar history and \textit{Insight}-HXMT observations allowed the
determination of the orbital ephemeris. The orbital period is 28 days, and the
eccentricity 0.1 \citep{doroshenko18,wilson18,zhang19}.

The X-ray spectrum, characterized by a cutoff power-law continuum
\citep{zhang19} and iron line emission \citep{jaisawal19}, is typical of
high-mass X-ray binaries. The strong magnetic field ($> 10^{12}$ G), 
estimated from accretion torque models and changes of the timing and the X-ray
spectral parameters during transitions between accretion regimes, confirms the
system as a hard X-ray transient \citep{doroshenko18,tsygankov18,doroshenko20}.
A tentative cyclotron line at $\sim10$ keV was reported by \citet{sugizaki20},
which would place the magnetic field of \swift\ in line with that of standard
accreting pulsars. One or more blackbody components associated with
the hot spot around the polar region and thermal emission from the accretion
column and from the photosphere of a possible ultra-fast outflow
\citep{tao19,eijnden19} complicate the spectrum at low energies.

The detection of radio emission
during the X-ray outburst is a surprising result \citep{eijnden18}. Radio emission in X-ray binaries
with no (black hole binaries) or weak (low-mass neutron star binaries) magnetic
field are common and attributed to the formation of a relativistic jet
\citep{fender04,migliari06}. However, it was believed that a strong magnetic
field, such as the field in accreting X-ray pulsars, inhibits jet formation. 
The detection of radio emission in \swift\ (a high-mass neutron star binary) has
also been interpreted as coming from a jet based on the correlation between
X-ray and radio luminosities, and on the radio spectral index evolution
\citep{eijnden18}. Nevertheless, the fact that the radio luminosity in \swift\
is two orders of magnitude fainter than seen in other neutron stars at similar
X-ray luminosities implies that the magnetic field still plays an important
role in regulating the jet power.

Several authors have derived different values for the
distance, from  $\sim2$ kpc based on optical photometric observations
\citep{bikmaev17} to $\sim 5-6$ kpc based on the spin-up rate and
accretion-torque models \citep{doroshenko18,zhang19}.  Although the distance
estimated from Gaia at face value is $\sim7$  kpc \citep{bailer-jones18}, a strong
lower limit of $d > 5$ kpc is found when all possible uncertainties in the Gaia analysis are taken into account \citep{eijnden18}. Even when this lower
limit is assumed for the distance, the peak of $1\times10^{39}$ \ergs\ implies that the
Eddington limit for the neutron star was exceeded during the outburst.

While there are numerous studies of \swift\ in the X-ray band
\citep{wilson18,doroshenko18,tsygankov18,jaisawal19,tao19,eijnden19,zhang19,doroshenko20,sugizaki20},
a detailed optical analysis is yet to be performed. A few days after its discovery
in the X-ray band, we carried out the first optical spectroscopic observations
from the 1.3 m telescope at the Skinakas Observatory \citep{kouroubatzakis17}. A
strong \ha\ line in emission and a blue-end spectrum that resembled that of an
early-type star led us to propose that \swift\ is a new Be/X-ray binary (BeXB). We here perform the first detailed analysis of the optical counterpart
to \swift\ and confirm this system as a new Be/X-ray binary.

\begin{table}[h]
\caption{Fast-photometry observations from Aras de los Olmos Observatory and average 
$B-$ and $V$-band magnitudes. Errors are 0.01 mag in both filters, except for MJD 58781, 
which is 0.02 mag.  $N_{\rm obs}$ is the number of observations in each filter.}
\label{fastlog}
\center
\begin{tabular}{ccccc}
\hline
\hline
MJD     &Date           &$B$    &$V$    & $N_{\rm obs}$  \\
\hline
\hline
58760   &03-10-2019     &13.81  &12.86   &43    \\
58765   &08-10-2019     &13.83  &12.87   &29    \\
58781   &24-10-2019     &13.78  &12.84   &29    \\
58782   &25-10-2019     &13.83  &12.87   &45    \\
58783   &26-10-2019     &13.82  &12.86   &37    \\
58883   &03-02-2020     &13.78  &12.83   &57    \\
58884   &04-02-2020     &13.78  &12.83   &22    \\
58885   &05-02-2020     &13.79  &12.84   &40    \\
58893   &13-02-2020     &13.77  &12.83   &32    \\
58894   &14-02-2020     &13.78  &12.83   &34    \\
58895   &15-02-2020     &13.79  &12.83   &23    \\
\hline
\end{tabular}
\end{table}
\begin{figure*}
\begin{center}
\resizebox{\hsize}{!}{\includegraphics{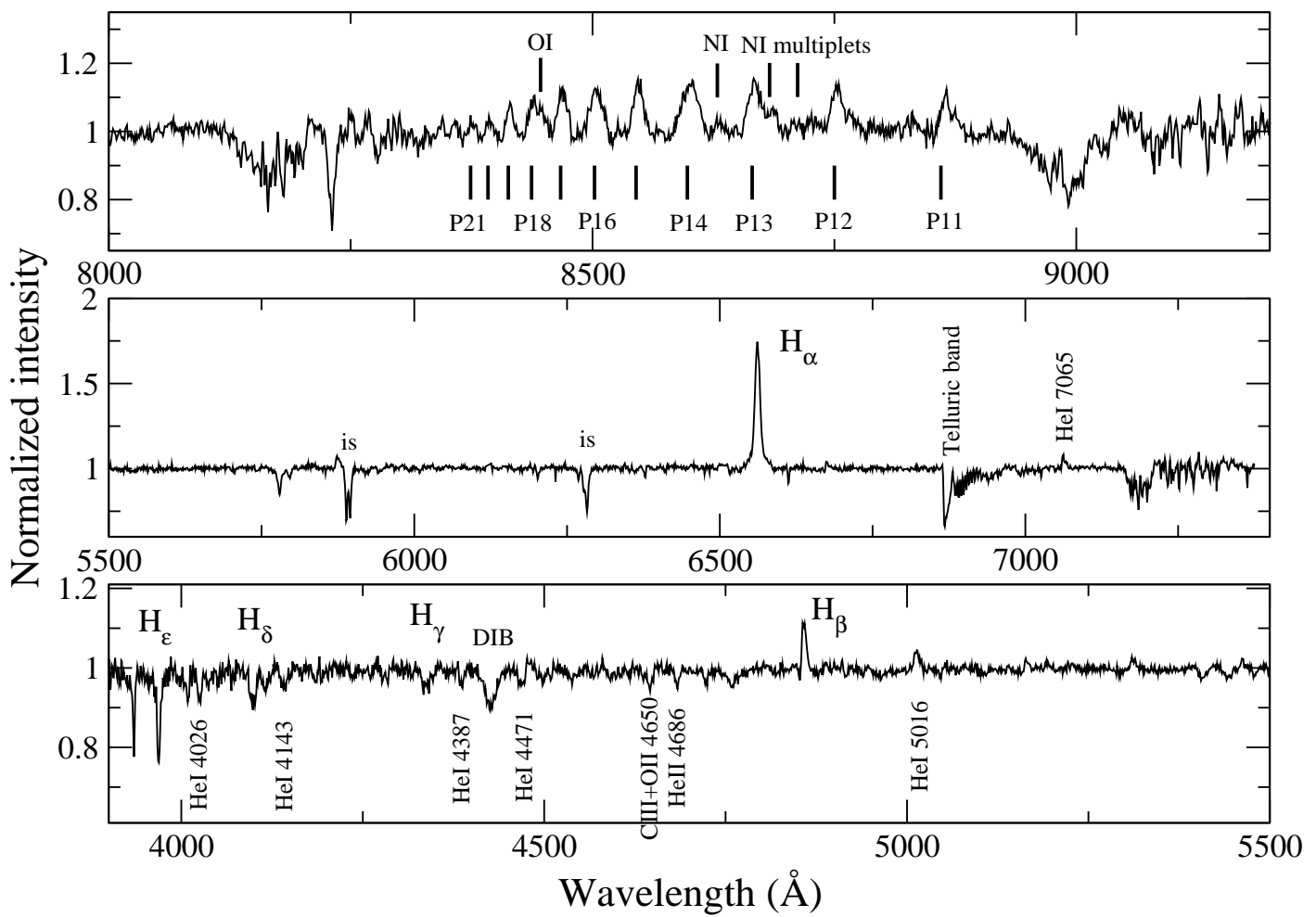}} 
\caption[]{Optical spectrum of \swift. The data were taken from the Skinakas observatory on 15 October 2017.}
\label{spec}
\end{center}
\end{figure*}

\begin{figure*}
\begin{center}
\includegraphics[width=16cm,height=11cm]{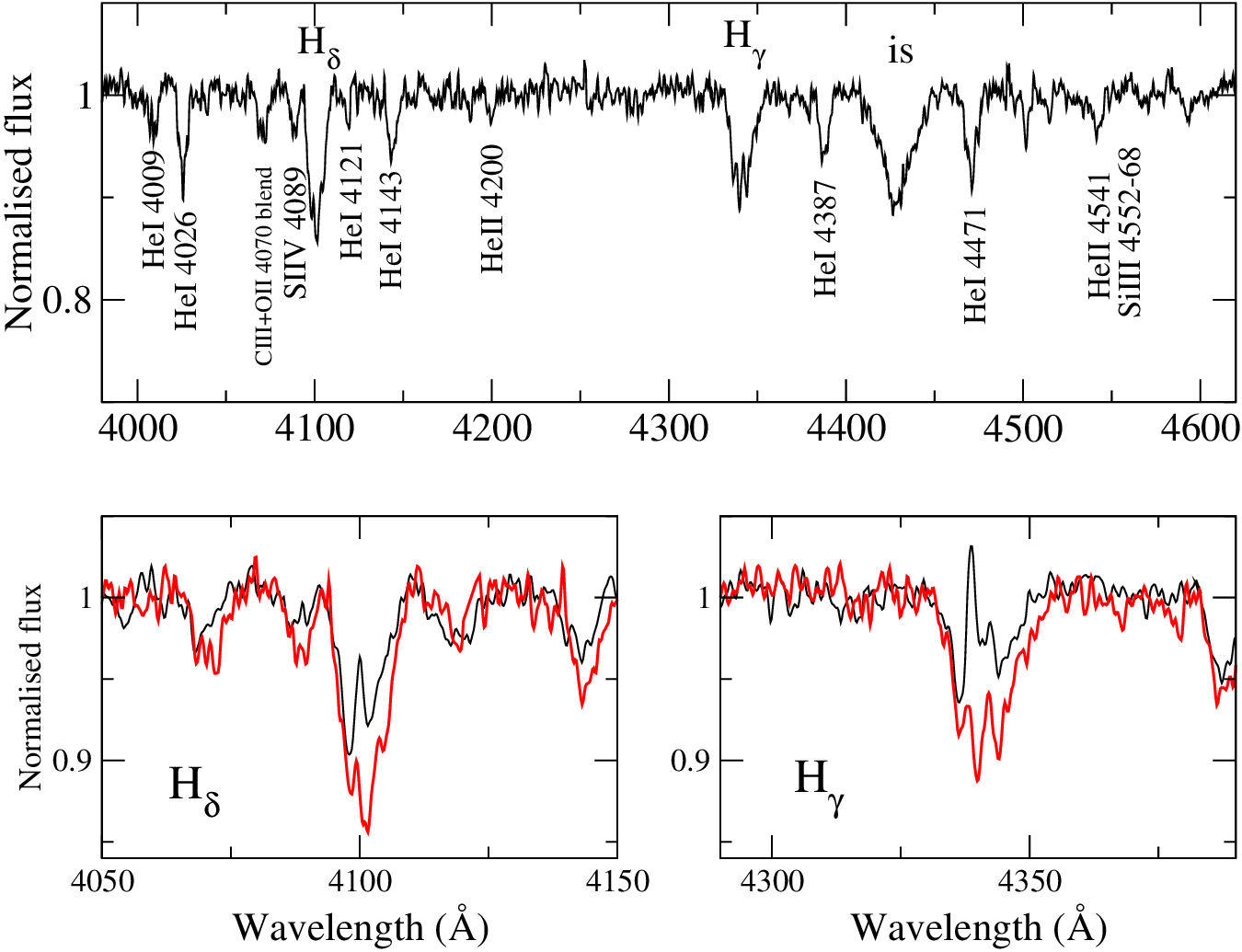} 
\caption[]{Optical spectrum showing the most relevant lines for spectral
classification. The spectrum in the top panel was obtained with the WHT 
(0.22 \AA/pixel) on 9 February 2020. The bottom panels display the spectrum 
taken with the WHT on 30 October 2017 (black), i.e., near the peak of the 
X-ray outburst, and on 9 February 2020 (red). The spectra were smoothed with a 
Gaussian filter with an FWHM of 0.7 \AA.}
\label{speclass}
\end{center}
\end{figure*}

\section{Observations}

\subsection{Spectroscopy}

Optical spectroscopic  observations were obtained from the 1.3m telescope of the
Skinakas observatory (SKO) in Crete (Greece). In addition, \swift\ was observed
in service time with the William Herschel Telescope (WHT) on the nights of 30
October 2017 and 9 February 2020.  The 1.3\,m telescope of the Skinakas
Observatory  was equipped with a 2048$\times$2048 (13.5 micron) pixels ANDOR
IKON CCD and a 1302 l~mm$^{-1}$ grating, giving a nominal dispersion of
$\sim$0.8 \AA/pixel.  The setup during the WHT observations was the ISIS
spectrograph with the Red+ and EEV12 CCDs for the red and blue arm,
respectively. The blue-sensitive EEV12 has an array of 4096$\times$2048 (13.5
micron) pixels, offering 0.23 \AA/pixel. The default detector for the ISIS red
arm is a red-sensitive array of 4096$\times$2048 (15.0 micron) pixels with
almost no fringing, giving 0.26 \AA/pixel.  Spectra of comparison lamps were
taken before each exposure in order to account for small variations of the
wavelength calibration during the night. To ensure an homogeneous processing of
the spectra, all of them were normalized with respect to the local continuum,
which  was rectified to unity by employing a spline fit. The log of the
spectroscopic observations is given in Table~\ref{speclog}.

\subsection{Photometry}

\subsubsection{Skinakas observatory}

Photometric observations with the Johnson-Cousins $B$,
$V$, $R,$ and $I$ filters were made with the 1.3m telescope of the
Skinakas Observatory. The telescope
was equipped with a  2048$\times$2048 ANDOR CCD with a 13.5 $\mu$m pixel
size. In this configuration, the  plate scale is 0.28$\arcsec$/pixel,
hence providing a field of view of $9.5 \times 9.5$ arcmin$^2$. Standard
stars from the Landolt list \citep{landolt09} were used for the
transformation equations.  The data were reduced in the
standard way using the IRAF tools for aperture photometry. After the
standardization process, we finally assigned an error to the calibrated
magnitudes of the target given by the rms of the residuals between the
cataloged and calculated magnitudes of the standard stars. The photometric
magnitudes are given in Table~\ref{photlog}.

\subsubsection{Aras de los Olmos observatory}

Photometric time series were obtained at the Aras de los Olmos Observatory
(OAO), located in La Muela de Santa Catalina, near of the Aras de los Olmos town
(Valencia, Spain), at an altitude of 1280 m. Observations were performed with
the 0.5m telescope equipped with a Finger Lakes Instruments ProLine PL16801
thermoelectrically cooled CCD camera. The 4096 $\times$ 4096 array with a 2
$\times$ 2 binning gives a plate scale of 1.08$\arcsec$/pixel and a field of
view of $36.9 \times 36.9$ arcmin$^2$. We performed differential photometry of
Swift\,J0243.6+6124 in the Johnson $B$ and $V$ band for five nights between 3 and 26
October 2019, and six more nights between 3 and 15 February 2020 
(Table~\ref{fastlog}). We transformed the differential instrumental magnitudes
into calibrated magnitudes using the measurements of the reference star GAIA
465628266540345216 ($\alpha$: 02 43 38.23, $\delta$: +61 26 40.7,  J2000)
taken from the Skinakas observatory. The average magnitudes of the reference
star are $B=13.67\pm0.01$,  $V=13.02\pm0.01$,  $R=12.65\pm0.01$, and 
$I=12.25\pm0.02$.

\subsubsection{ASAS-SN light curve}

We obtained the optical $V$--band light curve from the ASAS--SN Variable Stars
Database \citep{shappee14,jayasinghe19}. The pixel scale in ASAS--SN is
8\,arcsec, and the full width at half maximum (FWHM) is $\sim$2 pixels
\citep{jayasinghe19}. This means that blending has a considerable effect on the
data when a bright source lies within the FWHM. We used the results from the
Gaia Data Release 2 \citep{Evans2018} and found that there is a slightly fainter
star at 6.2\,arcsec from \swift. The calibrated $V$ -band magnitude of this
star, obtained from Skinakas data, is $V=14.52\pm0.01$ mag. We corrected our
observed magnitudes for contamination by removing the brightness of the close
star from the total observed flux because we considered that it is included in
the photometric aperture. The applied corrections slightly varied depending on
the variable flux of \swift, but they are in the range $\Delta
V$=0.18$\pm$0.01\,mag. Finally, we also used six observations from the AAVSO
obtained in the interval MJD 58532--58556.

\subsubsection{(Neo)WISE photometry}

We extracted the light curves in the $W1$ (3.4$\mu$m) and $W2$ (4.6$\mu$m) bands
provided by the Wide-field Infrared Survey Explorer
(\textit{WISE};\citealt{wright10}) and \textit{NeoWISE} \citep{mainzer11}
missions through the IRSA interface at \textit{https://irsa.ipac.caltech.edu}.

\section{Results}

\subsection{Spectral classification}

Figure~\ref{spec} shows the spectrum of \swift\ in the visible band. The
numerous He I and He II lines indicate that the optical companion to \swift\ is
an early-type star.  The emission lines of the Balmer and Paschen series imply
the presence of circumstellar material where stellar light is absorbed and
reemitted. These characteristics are typical of Be stars. This means that \swift\ is a
BeXB. To refine the spectral classification, we focus on the blue part of the
optical spectrum, in particular, on the 4000-4600 \AA\ region
(Fig.~\ref{speclass}),  and follow the atlas and method of
\citet{walborn90} and \citet{gray09}. In the bottom panels of
Fig.~\ref{speclass} we compare the two WHT spectra taken in October 2017 and
February 2020. The latest spectrum is less affected by disk emission, as is
apparent in the H$\gamma$ and H$\delta$ lines. 

The presence of He I and He II lines, moderate Si IV 4089, the absence of Mg II
4481, and the weakness of Si III 4552-68 indicate a type earlier than B0.5 
Although He II 4686 and He II 4541 are strong, the ratio between  He I 4387 and
He II 4541 is clearly greater than 1, which implies a O9.5 star. This ratio is
$\sim$ 1 for spectral type O9 and reverses for hotter stars. Likewise, the
comparable intensity of He II 4686 and the C III+O II blend at 4650 \AA\
(Fig.~\ref{spec}) together with the weak He II 4200 rules out a type earlier
than O9. For example, an O8--O8.5V would have  He II 4686 $>>$ CIII+OII as well
as He II 4200 $>>$ He I 4121-4143.

Some traces of Si IV 4116 may be present but very weak. The ratio Si IV 4116/He
I 4121 $<<$1 favors a O9-B0. It also suggests a luminosity class V because in
giants and supergiants, this ratio is $>> 1$. The strength of the He II
lines also indicates a luminosity class V.   The
absorption strength of this line weakens in giants and supergiants. 

In the red part of the spectrum,   from P11 to P21, together with the conspicuous Paschen lines
series, the  O I 8446 line is in emission and blended with
P18. The N I 8680-8683-8686 multiplet, blended in a single broad feature, also
appears in emission near the right wing of the P13 line. The other N I features
in this spectral range, the 8629 \AA\ line and the 8703-8712-8719 multiplet, are
marginally detected in emission.  When these lines are detected in Be stars, they are
always present in emission in early spectral subtypes up to B2.5
\citep{andrillat88}. 
We conclude that the most likely spectral type of the companion of \swift\ is
O9.5Ve.

\subsection{Reddening and distance}

To estimate the distance through the distance-modulus relation, $V-M_V-A_V=5
\log(d)-5$, the amount of interstellar extinction $A_{\rm V}=R \times E(B-V)$ to
the source has to be determined.  We used two methods to derive the color
excess $E(B-V)$. The most direct method is to use the calibrated color of the star according
to the spectral type. The color excess is defined as $E(B-V)= (B-V)_{\rm
obs}-(B-V)_0$, where $(B-V)_{\rm obs}$ is the observed color and $(B-V)_0$ is
the intrinsic color of the star. The second method uses the strength of certain
interstellar lines to estimate the interstellar reddening
\citep{herbig75,herbig91,galazutdinov00,friedman11,kos13}. 

The 2019 photometric observations
gave $(B-V)_{\rm obs}=0.95\pm0.02$ in July and $(B-V)_{\rm obs}=0.97\pm0.02$ in
September. The observed color depends on the spectral type of the star. Based on
our derivation of the spectral type, the expected color for an O9.5V star is
$(B-V)_0=-0.29\pm0.02$. This value is the average of the calibrations from
\citet{johnson66,fitzgerald70,gutierrez79,wegner94,pecaut13} and 
the error is the standard deviation of the five values.  The color excess is
then estimated to be $E(B-V)=1.24\pm0.02$.

The color excess can also be estimated from the strength of the diffuse
interstellar bands  (DIB). When the strongest and cleanest lines (5780\AA\
and 6613\AA) in the red spectrum of \swift and the calibrations of
\citet{herbig75} and \citet{friedman11} are used, the estimated color excess is
$E(B-V)=1.1\pm0.2$, in agreement with the photometric derived value.  The error
is the standard deviation of all measurements.    

When the standard extinction law $R=3.1$ is used, the extinction in the $V$ band is
then $A_V=R \times E(B-V)=3.41-3.84$ mag. This lies in between the extinction along
the line of sight of \swift\ given by the reddening analysis of
\citet{schlafly11} ($A_V=3.48$ mag) and \citet{schlegel98} ($A_V=4.19$
mag)\footnote{Obtained using the Nasa Extragalactic Database extinction
calculator (https://ned.ipac.caltech.edu/extinction\_calculator).}.

With the photometrically derived value of the color excess, $A_V=3.84$,
$V=12.90\pm0.02$, and an absolute magnitude for an O9.5V star of $M_{\rm
V}=-4.2$, obtained as the average  from the calibrations of
\citet{humphreys84,vacca96,martins05,wegner06,pecaut13}, the distance to \swift\
is estimated to be  $4.5\pm0.5$ kpc. 

There are a number of uncertainties in the derivation of the distance using
photometry that are related to the uncertainty in the calibrations of the
absolute magnitude and colors associated with the spectral type and luminosity
class. Differences of 0.04 mag in intrinsic $(B-V)_0$ are found in studies from
different authors. Likewise, the relation between absolute magnitude and
luminosity has a large intrinsic dispersion \citep{jaschek98,wegner06}.   The
error in the distance was obtained by propagating the errors 0.02 mag in $V$
and $E(B-V)$ and 0.2 mag in $M_{\rm V}$.

The optical companion to  \swift\ is an OBe star. This adds an additional
source of uncertainty in the determination of the distance. The circumstellar disk means that the photometric magnitudes and colors may be
contaminated by disk emission.  The overall effect is to cause the star to appear
redder, that is, to be seen at higher extinction \citep{fabregat90,riquelme12}.
As a result, the distance estimate that is obtained from photometric magnitudes
that may be to be affected by disk emission should be taken as a lower limit.  The
photometric observations that we used to derive the distance of $\sim$4.5 kpc
corresponded to the lowest value of the equivalent width (EW) \ew, that is, when the contribution of
the disk was expected to be minimum. Some residual emission is still present
(Table~\ref{speclog}). When the additional reddening from the disk is taken into account
and Eq. (21) from \citet{riquelme12} is applied, the distance increases to
$4.8\pm0.5$ kpc. The value of the \ha\ EW used in this correction
is derived from the EW of the measured spectrum (-5.8 \AA, Table~\ref{speclog})
and the fill-in emission for a O9.5V star \cite[see, e.g., Table 4
in][]{riquelme12}.

When we instead use the color excess from the DIB analysis, we obtain a distance
of $5.5\pm1.7$ kpc. Here the main source of error lies in the uncertainty of
$E(B-V)$.

Our value of the distance is similar to that obtained from an X-ray analysis of
the correlation between the spin-up rate and the magnetic field of the neutron
star of $\sim 5$ kpc \citep{doroshenko18}. Based on {\it Gaia} observations,
\citet{eijnden18} set a lower limit of 5 kpc, while the {\it Gaia}
catalog  \citep{bailer-jones18} gives $6.8^{+1.5}_{-1.1}$ kpc.

\begin{table*}
\begin{center}
\caption{Comparison of \swift\ with other Be/X-ray binaries. }
\label{rotvel}
\begin{tabular}{lllcccc}
\noalign{\smallskip} \hline \noalign{\smallskip}
X-ray           &Optical        &Spectral&Disk-loss     &P$_{\rm orb}$  &$v \sin i$ &Reference \\
source          &counterpart    &type   &episodes       &(days)         &(km s$^{-1}$)& \\
\noalign{\smallskip} \hline \noalign{\smallskip}
\swift\         &--             &O9.5V  &no     &27.8      &210$\pm$20     &This work \\
4U 0115+634     &V635 Cas       &B0.2V  &yes    &24.3      &300$\pm$50     &1       \\
RX J0146.9+6121 &LS I +61 235   &B1V    &no     &--        &200$\pm$30     &2       \\
V 0332+53       &BQ Cam         &O8-9V  &no     &34.2      &$<$150         &3       \\
X-Per           &HD 24534       &O9.5III&yes    &250       &215$\pm$10     &4,5    \\
RX J0440.9+4431 &LS V +44 17    &B1III-V&yes    &150       &235$\pm$15     &6,7    \\
1A 0535+262     &HD 245770      &O9.7III&yes    &111       &225$\pm$10     &8,9    \\
IGR J06074+2205 &--             &B0.5IV &yes    &--        &260$\pm$20     &10     \\
RX J0812.4-3114 &LS 992         &B0.5III-V&yes  &81.3      &240$\pm$20     &11     \\
1A 1118-615     &Hen 3-640      &O9.5IV &no     &24        &$\sim$300      &12,13  \\
4U 1145-619     &V801 Cen       &B0.2III&no     &187       &280$\pm$30     &14,15  \\
4U 1258-61      &V850 Cen       &B2V    &yes    &132       &$<$600  &16     \\
SAX\,J2103.5+4545&--            &B0V    &yes    &12.7      &240$\pm$20     &17     \\
IGR\,J21343+4738&--             &B1IV   &yes    &--        &365$\pm$15  &18   \\
SAX\,2239.3+6116&--             &B0V    &no     &262.6     &195$\pm$20     &19     \\
\noalign{\smallskip} \hline
\end{tabular}
\end{center}
\tablebib{(1) \citet{negueruela01a}; (2) \citet{reig97b}; (3) \citet{negueruela99}; (4) \citet{lyubimkov97};
(5) \citet{delgado01}; (6) \citet{reig05b}; (7) \citet{ferrigno13}; (8) \citet{haigh04};
(9) \citet{grundstrom07b}; (10) \citet{reig10b}; (11) \citet{reig01}; (12) \citet{janot81};
(13) \citet{staubert11}; (14) \citet{janot82}; (15) \citet{alfonso17}; (16) \citet{parkes80};
(17) \citet{reig04}; (18) \citet{reig14a}; (19) \citet{reig17}
}
\end{table*}

\subsection{Rotational velocity}

The rotational velocity is an important parameter that it is linked to the
formation of the circumstellar disk and allows us to estimate the velocity law of
the gas particles in the disk \citep[see, e.g., the discussion in][]{reig16}.
Because Be stars are fast rotators that likely spin at close to the critical
rotation rate at which centrifugal forces balance Newtonian gravity,  the
effective equatorial gravity is reduced to the extent that weak processes such as
gas pressure and/or nonradial pulsations alone or in combination with magnetic
flaring activity  may trigger the ejection of photospheric matter with
sufficient energy and angular momentum to form a Keplerian disk.

The use of optical spectra to estimate the rotational velocity is based on the
fact that stellar absorption lines are rotationally broadened. Thus the
projected rotational velocity, $v \sin i$, where $v$ is the equatorial
rotational velocity and $i$ is the inclination angle toward the observer, 
can be estimated by measuring the width of lines that are not affected by the disk
contribution. \ion{HeI}\ lines are significantly less affected by disk emission
than hydrogen lines and are considerably stronger than other metallic lines.
Thus we estimated the projected rotational velocity by measuring the FWHM of the \ion{HeI}\ lines.

For this analysis, we used the WHT spectra taken on the night of 9 February 2020 as it
provides the highest spectral resolution and the least contamination from disk
emission (see Fig.~\ref{speclass}). Nevertheless, we note that some level of the
disk emission remains in H and He lines, as implied by the fact that the \ha\
line showed an emission profile with \ew=--7.5 \AA. Residual emission has the
effect to increase the width of the line because the FWHM is measured not at
half maximum (or minimum for an absorption line), but close to the continuum
where the line profile is broader.

We fit a Gaussian profile and derived the FWHM of four \ion{HeI}\ lines (4026
\AA, 4143 \AA, 4387 \AA, and 4471 \AA) on each one of the two spectra taken by
WHT on the night of 9 February 2020 as well as on the average. We repeated the
procedure three times, which corresponds to different selections of the continuum.
The final rotational velocity is the mean of all these measurements. The
conversion from the FWHM into velocity was made using the calibration by
\citet{steele99}, after correcting for instrumental broadening ($FWHM_{\rm
ins}=55$ \kms). The average and standard deviation gave $v \sin i = 210\pm20$ km
s$^{-1}$. Table~\ref{rotvel} compares the rotational velocity of \swift\ with
other BeXBs.

\subsection{Fast photometry. Frequency analysis}

We performed a frequency analysis of the Aras de los Olmos differential
photometry data by means of the standard Fourier analysis techniques and
least-squares fitting using the codes {\tt Period04} \citep{lenz05} and PASPER
\citep{diago08}. These programs determine the frequency with the largest amplitude by
performing a Fourier transform and then fitting the light curve with a
sinusoidal function using a least-squares algorithm. Then this frequency is
removed and a new search is performed. The method is iterative and ends when the
removal of a new frequency is not statistically significant. 

As a result of the analysis no significant frequency was found in the interval of 0
to 20  cycles day$^{-1}$. The accuracy of the differential photometry,
which is estimated as the standard deviation of the mean differential photometry between
nonvariable reference stars in the field, amounts to 7 mmag in the $B$
and $V$ bands. We therefore conclude that no short-term variability in the amplitude
larger than 7 mmag and frequency lower than 20  cycles day$^{-1}$ was
present in \swift\ during the time span of our observations.

\begin{figure} 
\resizebox{\hsize}{!}{\includegraphics{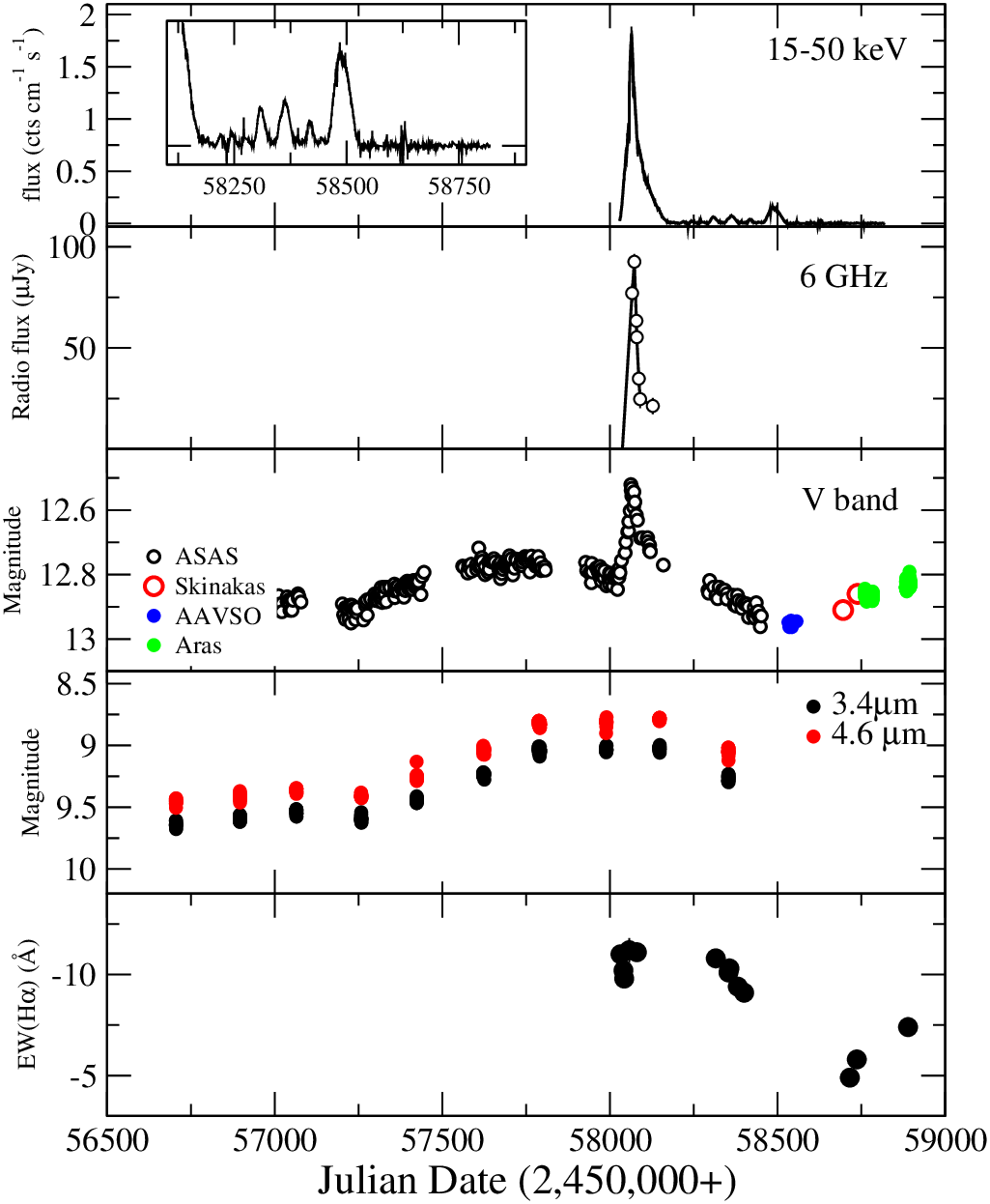}} 
\caption[]{Multiwavelength variability of  \swift: {\it Swift}/BAT X-ray flux,
radio flux \citep[from][]{eijnden18}, V-band
magnitude, NEOWISE infrared magnitudes at 3.4 $\mu$m and 4.6 $\mu$m, and
\ha\ EW. The inset in the
top panel shows the series of type I outbursts that followed the main giant
outburst.}
 \label{xoptirad}
\end{figure}

\section{Discussion}

The optical emission in BeXBs mainly comes from two different regions: the
circumstellar disk, and the star itself. Long-term variations (on the order of
months to years) are related to structural changes in the disk, while fast
variability ($\text{of about some}$ hours) is attributed to changes in the stellar photosphere.
The nondetection of fast photometric variability in \swift\ represents an
additional peculiarity with respect to the common characteristics of the BeXB
class. Virtually all BeXBs that are searched for short-term variability display
intra-night photometric modulations in their light curves with the same range of
frequencies and amplitudes as analyzed for \swift\ (Reig et al., in
preparation).

Figure~\ref{xoptirad} shows the long-term evolution of the optical parameters
with time. The giant outburst that led to the discovery of the source as an 
X-ray transient in October 2017 is clearly seen in the top panel of this figure.
The brightening during the years prior to the X-ray outburst and the subsequent
decrease in the optical and infrared emission agree with a disk origin and is
typical of BeXBs \citep{reig16}. Be star disks are known to form and dissipate
on timescales of years. The disk emission increases with wavelength and becomes
particularly significant in the infrared band. The optical and infrared light
curves in Fig.~\ref{xoptirad} indicate that the disk grew steadily during the years prior to the X-ray
outburst. The optical and infrared emission reached a
local maximum in the weeks before the X-ray outburst occurred, clearly
indicating that a large disk was already present by the time the X-ray
outburst was detected. The disk radius most likely extended all the way up to
the periastron distance.

In about October 2017, the neutron star
interacted with the disk, accreting a large amount of material and producing the
large X-ray flare. After the main X-ray outburst, a series of type I outbursts
followed.  As a result of the X-ray activity, the disk weakened. This became
evident by the significant decrease in the EW of the \ha\ line
from a maximum of --11 \AA\ during the peak of the X-ray outburst (MJD 58031) to
a minimum of --4.9 \AA\ (MJD 58715) two years later.
Further evidence for a smaller disk during the 2020 observation comes from the
significantly lower fill-in emission that affects the Balmer lines compared to the
October 2017 spectrum (see the bottom panels in Fig.~\ref{speclass}).

The variability of the optical emission during the X-ray outburst is
unusual among BeXBs, however. The optical and X-ray luminosity are strongly correlated
 without any significant delay.
Smooth and slow variations in the optical light curve of Be stars in BeXB of a
few tenths of magnitude are common and attributed to the disk formation and
dissipation \citep{reig15}. The optical and IR brightening
of the Be star normally precedes the appearance of X-ray outbursts in BeXBs
\citep{stevens97,coe06,reig07b,camero12,camero14,caballero16,alfonso17}.
Although long-term photometric  coverage with good cadence is scarce for
BeXBs, the available data show that this long-term variability is gradual and
slow and contrasts with the sudden increase and subsequent decrease that is observed
during the X-ray outburst in \swift. The tight correlation between the X-ray,
radio, and $V$-band emission during the X-ray outburst suggests a strong link
between the processes that give rise to the emission at the different wavelength
bands. A detailed study of the optical outburst and its
correlation with the X-ray flux will be presented in a forthcoming paper
(Alfonso-Garz\'on et al. 2020, in preparation).

\begin{figure}
\begin{center}
\includegraphics[width=8cm]{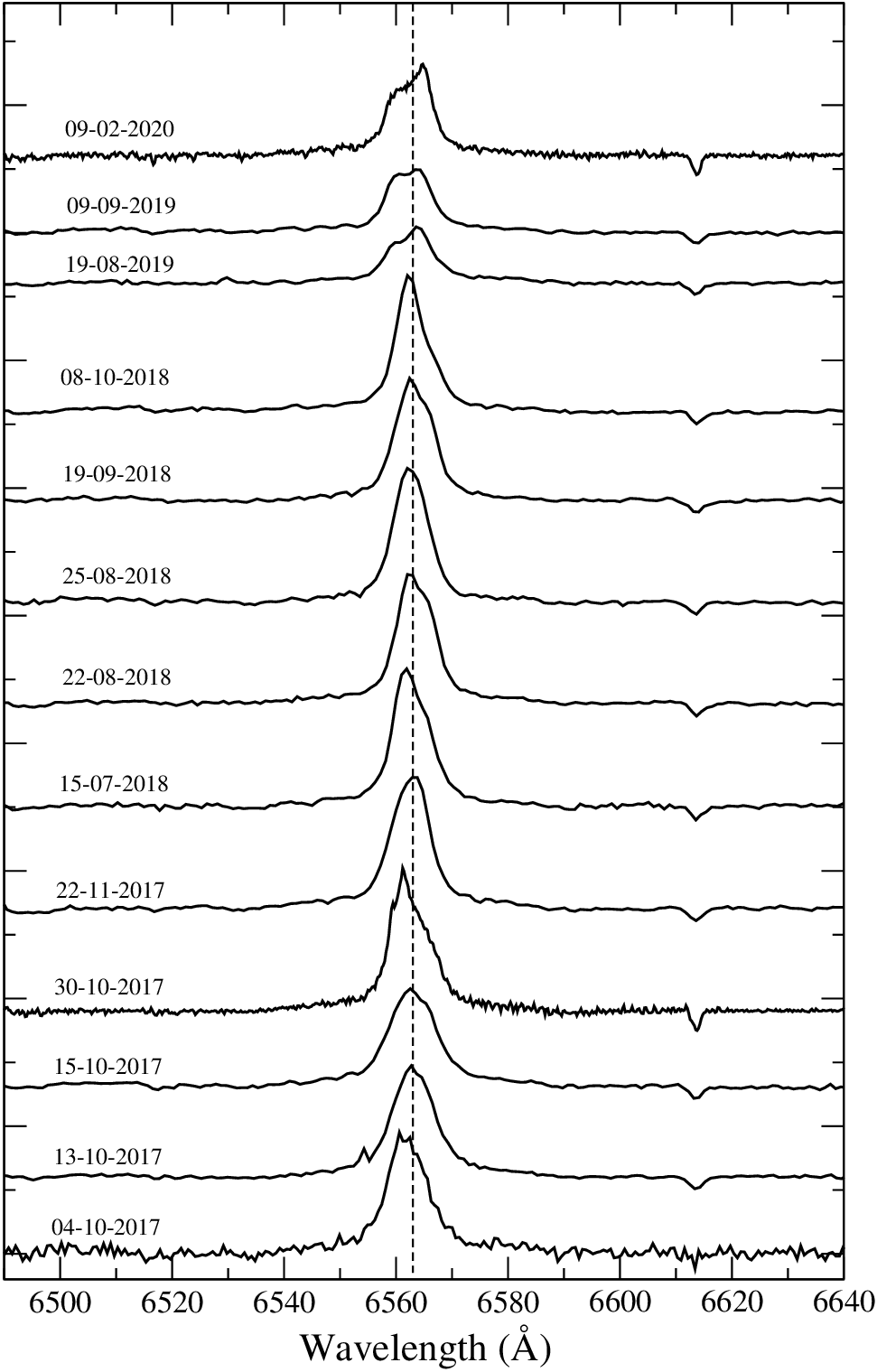} 
\caption[]{Evolution of the \ha\ line profile. The vertical line marks the
rest wavelength of the \ha\ line.}
\label{lineprof}
\end{center}
\end{figure}

The profile of the \ha\ line provides further information about the state of the
circumstellar disk. Be stars show a wide range of different shapes. These shapes are
attributed to different inclination angles  \citep[][but see also
\citealt{silaj10}]{slettebak79,rivinius13a}. At low inclination, the line
typically shows a single-peak profile. At intermediate angles, Be stars exhibit
a double-peak profile. The double-peak structure is a natural consequence of
rotation velocity prevailing over radial displacements. At high inclination, the
double-peak presents a central depression caused by self absorption from the
disk (shell profile).

Unfortunately, the low spectral resolution of our observations does not allow a
detailed study of the changes in the shape of the line, but the data show that
\swift\ displayed a single-peaked profile that turned into a double-peaked
profile at low values of \ew\ (Fig.~\ref{lineprof}). This is the expected
behavior in Keplerian disks, where the rotational velocity of the gas particles
in the disk varies as $v_{\phi}\propto r^{-1/2}$. Because the  peak separation in
a double-peaked profile is directly linked with this velocity, the peaks separation increases when the disk radius decreases. Nevertheless, even with the
high spectral resolution provided by the WHT spectrum, the double peak is not
clearly distinguishable.  This suggests that we see the circumstellar disk
at a small inclination angle and that it resembles the behavior of V\,0332+54, which
always displays a single-peak line \citep{reig16,caballero16}. The
inclination of V 0332+54 is estimated to be $i<20^{\circ}$ \citep{negueruela99,zhang05}.
Further evidence for a low-inclination angle of the circumstellar disk is given
by the WHT spectrum, which has a higher resolution and was obtained near the peak of the outburst. This spectrum displays inflections in the flanks of the profile that are reminiscent of a
wine-bottle profile. This type of profile results from incoherent scattering
broadening that is caused by the optical thickness of the \ha\ line in the direction
perpendicular to the disk plane and appears at small inclination angles
\citep{hanuschik86,hummel95,hummel97}.

On the other hand, we can roughly estimate the inclination of the orbital plane
from the orbital solution. The projected semimajor axis is $a \sin i \sim 116$
lt-s (or $\sim50$ $\rsun$) and the orbital period is 27.8 d
\citep{wilson18,zhang19}. When we assume typical masses for the O9.5V star and
neutron star of 16 $\msun$ and 1.4 $\msun$, we obtain from the third Kepler
law $a\sim 100$ $\rsun$. Hence $i\sim 30^{\circ}$.

\subsection{Giant X-ray outburst}

Current models that explain giant (type II) X-ray outbursts in BeXBs are based
on the idea that these outbursts occur when the neutron star captures a large amount of gas
from a warped \citep{martin11,okazaki13,moritani13} or a highly eccentric disk
\citep{martin14a}. The warping of the disk may be caused by the tidal
interaction with the neutron star \citep{martin11} or by radiation from the
central star \citep{porter98}, whereas the main mechanism that has been
identified to produce eccentricity growth is Kozai-Lidov oscillations
\citep[][but see also \citealt{martin14a}]{martin19}. In either case, the
unavoidable condition for these two processes (warping and eccentricity growth)
to work is the misalignment  of the decretion disk. In other words, the disk
must be tilted. However, the two mechanisms differ in the value of the initial
tilt angle, that is, the angle between the spin axis of the Be star and that of the
binary orbit, or angle between the orbital plane and the Be disk. While the
tidal torque scenario favors small tilt angles, $\beta < 20^{\circ}$
\citep{okazaki13}, the eccentric disk scenario is expected only in systems with
a large misalignment $\beta > 60^{\circ}$ \citep{martin14a}.
From the point of view of the observations, observational evidence for
misaligned disks during X-ray outbursts comes from the variability in the shape
of the spectral emission lines \citep{negueruela01b,moritani11,moritani13} and
changes in the polarization angle \citep{reig18b}. 

Although we do not have optical spectra prior to the 2017 X-ray outburst that
would allow us to gauge possible inclination effects from changes in the line
profile, the data provide some evidence for a small tilt angle. We estimated the
orbital inclination to be $\sim30^{\circ}$, while we argued that the single peak
and inflections in the flanks of the \ha\ line (wine-bottle profile) also favor
a small disk inclination.   Indeed, \swift\ shows some elements that could be
the result of a warped disk, or more generally, of an asymmetrical distribution
of matter in the disk. The X-ray activity of \swift\ continued after the major
outburst for about a year in the form of minor orbitally modulated ($P_{\rm
orb}=28$ d) outbursts. The peak intensity of these minor outbursts was distributed
irregularly, with values in the range $\sim0.02-0.2$  BAT counts cm$^{-2}$
s$^{-1}$, without any  ordered time sequence. This irregularity could be
interpreted as the neutron star probing different parts of the disk with
different density at each periastron passage as a result of a warped disk.  The
optical emission also shows some features that could be attributed to an
inhomogeneous disk (see Fig.~\ref{xoptirad}): the local minimum at $\sim$MJD
58000 (about one month before the onset of the outburst) in the $V$-band light
curve, the sudden decrease  in the EW of the \ha\ line by
$\sim$10\% during the rise at MJD 58040, and the rebrightening in $V$ mag during
the decay at around MJD 58100.

After the outburst, the outer parts of the disk disappeared because they
were accreted onto the neutron star. The disk, however, did not seem to return to a normal quiescent state until after the end of the X-ray activity at
around MJD 58550.

\section{Conclusion}

The long-term optical variability of \swift\ is typical
of BeXBs. The disk grew until its radius became about as large as the periastron distance. This growth led to
mass transfer onto the neutron star and caused a giant X-ray outburst. After
the outburst, the decrease in optical continuum  emission and in the strength
of the \ha\ line indicated that the disk weakened. However,  a disk-loss episode
never occurred. The latest observations suggest that a new growth
phase has begun.   We speculate that a warped disk is responsible for the X-ray variability that was observed
during and after the main outburst. While the long-term optical variability does not reveal any peculiarity that could explain the unique nature of \swift, the lack of fast photometric variability driven by nonradial pulsations is surprising.

\begin{acknowledgements}

Skinakas Observatory is run by the University of Crete and the Foundation for
Research and Technology-Hellas.  The Aras de los Olmos Observatory (OAO) is a facility of the Astronomical Observatory of the Valencia University (Spain). We thank O. Brevi\'a and V. Peris for their support to the OAO observations. This proposal benefited by the WHT Service
proposals SW2017b10 and SW2019b18. The WHT and its service program are operated
on the island of La Palma by the Isaac Newton Group of Telescopes in the Spanish
Observatorio del Roque de los Muchachos of the Instituto de Astrof\'{\i}sica de
Canarias. This publication makes use of data products from the Wide-field
Infrared Survey Explorer, which is a joint project of the University of
California, Los Angeles, and the Jet Propulsion Laboratory/California Institute
of Technology, funded by the National Aeronautics and Space Administration. This
publication also makes use of data products from NEOWISE, which is a project of
the Jet Propulsion Laboratory/California Institute of Technology, funded by the
Planetary Science Division of the National Aeronautics and Space Administration.

\end{acknowledgements}

\bibliographystyle{aa}
\bibliography{../../artBex_bib}

\begin{thebibliography}{92}
\expandafter\ifx\csname natexlab\endcsname\relax\def\natexlab#1{#1}\fi

\bibitem[{{Alfonso-Garz{\'o}n} {et~al.}(2017){Alfonso-Garz{\'o}n}, {Fabregat},
  {Reig}, {Kajava}, {S{\'a}nchez-Fern{\'a}ndez}, {Townsend}, {Mas-Hesse},
  {Crawford}, {Kretschmar}, \& {Coe}}]{alfonso17}
{Alfonso-Garz{\'o}n}, J., {Fabregat}, J., {Reig}, P., {et~al.} 2017, \aap, 607,
  A52

\bibitem[{{Andrillat} {et~al.}(1988){Andrillat}, {Jaschek}, \&
  {Jaschek}}]{andrillat88}
{Andrillat}, Y., {Jaschek}, M., \& {Jaschek}, C. 1988, \aaps, 72, 129

\bibitem[{{Bahramian} {et~al.}(2017){Bahramian}, {Kennea}, \&
  {Shaw}}]{bahramian17}
{Bahramian}, A., {Kennea}, J.~A., \& {Shaw}, A.~W. 2017, The Astronomer's
  Telegram, 10866, 1

\bibitem[{{Bailer-Jones} {et~al.}(2018){Bailer-Jones}, {Rybizki}, {Fouesneau},
  {Mantelet}, \& {Andrae}}]{bailer-jones18}
{Bailer-Jones}, C.~A.~L., {Rybizki}, J., {Fouesneau}, M., {Mantelet}, G., \&
  {Andrae}, R. 2018, \aj, 156, 58

\bibitem[{{Bikmaev} {et~al.}(2017){Bikmaev}, {Shimansky}, {Irtuganov},
  {Glushkov}, {Sakhibullin}, {Khamitov}, {Burenin}, {Lutovinov}, {Zaznobin},
  {Pavlinsky}, {Sunyaev}, {Dodonov}, {Afanasiev}, {Kotov}, {Doroshenko}, \&
  {Tsygankov}}]{bikmaev17}
{Bikmaev}, I., {Shimansky}, V., {Irtuganov}, E., {et~al.} 2017, The
  Astronomer's Telegram, 10968

\bibitem[{{Caballero-Garc{\'\i}a} {et~al.}(2016){Caballero-Garc{\'\i}a},
  {Camero-Arranz}, {{\"O}zbey Arabac{\i}}, {Zurita}, {Suso},
  {Guti{\'e}rrez-Soto}, {Beklen}, {Kiaeerad}, {Garrido}, \&
  {Hudec}}]{caballero16}
{Caballero-Garc{\'\i}a}, M.~D., {Camero-Arranz}, A., {{\"O}zbey Arabac{\i}},
  M., {et~al.} 2016, \aap, 589, A9

\bibitem[{{Camero} {et~al.}(2014){Camero}, {Zurita}, {Guti{\'e}rrez-Soto},
  {{\"O}zbey Arabac{\i}}, {Nespoli}, {Kiaeerad}, {Beklen}, {Garc{\'\i}a-Rojas},
  \& {Caballero-Garc{\'\i}a}}]{camero14}
{Camero}, A., {Zurita}, C., {Guti{\'e}rrez-Soto}, J., {et~al.} 2014, \aap, 568,
  A115

\bibitem[{{Camero-Arranz} {et~al.}(2012){Camero-Arranz}, {Finger},
  {Wilson-Hodge}, {Jenke}, {Steele}, {Coe}, {Gutierrez-Soto}, {Kretschmar},
  {Caballero}, {Yan}, {Rodr{\'\i}guez}, {Suso}, {Case}, {Cherry}, {Guiriec}, \&
  {McBride}}]{camero12}
{Camero-Arranz}, A., {Finger}, M.~H., {Wilson-Hodge}, C.~A., {et~al.} 2012,
  \apj, 754, 20

\bibitem[{{Coe} {et~al.}(2006){Coe}, {Reig}, {McBride}, {Galache}, \&
  {Fabregat}}]{coe06}
{Coe}, M.~J., {Reig}, P., {McBride}, V.~A., {Galache}, J.~L., \& {Fabregat}, J.
  2006, \mnras, 368, 447

\bibitem[{{Delgado-Mart{\'{\i}}} {et~al.}(2001){Delgado-Mart{\'{\i}}},
  {Levine}, {Pfahl}, \& {Rappaport}}]{delgado01}
{Delgado-Mart{\'{\i}}}, H., {Levine}, A.~M., {Pfahl}, E., \& {Rappaport}, S.~A.
  2001, \apj, 546, 455

\bibitem[{{Diago} {et~al.}(2008){Diago}, {Guti{\'e}rrez-Soto}, {Fabregat}, \&
  {Martayan}}]{diago08}
{Diago}, P.~D., {Guti{\'e}rrez-Soto}, J., {Fabregat}, J., \& {Martayan}, C.
  2008, \aap, 480, 179

\bibitem[{{Doroshenko} {et~al.}(2018){Doroshenko}, {Tsygankov}, \&
  {Santangelo}}]{doroshenko18}
{Doroshenko}, V., {Tsygankov}, S., \& {Santangelo}, A. 2018, \aap, 613, A19

\bibitem[{{Doroshenko} {et~al.}(2020){Doroshenko}, {Zhang}, {Santangelo}, {Ji},
  {Tsygankov}, {Mushtukov}, {Qu}, {Zhang}, {Ge}, {Chen}, {Bu}, {Cao}, {Chang},
  {Chen}, {Chen}, {Chen}, {Chen}, {Chen}, {Cui}, {Cui}, {Deng}, {Dong}, {Du},
  {Fu}, {Gao}, {Gao}, {Gao}, {Gu}, {Guan}, {Guo}, {Han}, {Hu}, {Huang}, {Huo},
  {Jia}, {Jiang}, {Jiang}, {Jin}, {Jin}, {Kong}, {Li}, {Li}, {Li}, {Li}, {Li},
  {Li}, {Li}, {Li}, {Li}, {Li}, {Li}, {Li}, {Liang}, {Liao}, {Liu}, {Liu},
  {Liu}, {Liu}, {Liu}, {Liu}, {Liu}, {Lu}, {Lu}, {Lu}, {Luo}, {Ma}, {Meng},
  {Nang}, {Nie}, {Ou}, {Sai}, {Shang}, {Song}, {Song}, {Sun}, {Tan}, {Tao},
  {Tuo}, {Wang}, {Wang}, {Wang}, {Wang}, {Wen}, {Wu}, {Wu}, {Xiao}, {Xiong},
  {Xu}, {Xu}, {Yang}, {Yang}, {Yang}, {Yang}, {Zhang}, {Zhang}, {Zhang},
  {Zhang}, {Zhang}, {Zhang}, {Zhang}, {Zhang}, {Zhang}, {Zhang}, {Zhang},
  {Zhang}, {Zhang}, {Zhang}, {Zhang}, {Zhang}, {Zhang}, {Zhao}, {Zhao}, {Zhao},
  {Zheng}, {Zhu}, {Zhu}, {Zou}, \& {Zhang}}]{doroshenko20}
{Doroshenko}, V., {Zhang}, S.~N., {Santangelo}, A., {et~al.} 2020, \mnras, 491,
  1857

\bibitem[{{Evans} {et~al.}(2018){Evans}, {Riello}, {De Angeli}, {Carrasco},
  {Montegriffo}, {Fabricius}, {Jordi}, {Palaversa}, {Diener}, {Busso},
  {Cacciari}, {van Leeuwen}, {Burgess}, {Davidson}, {Harrison}, {Hodgkin},
  {Pancino}, {Richards}, {Altavilla}, {Balaguer-N{\'u}{\~n}ez}, {Barstow},
  {Bellazzini}, {Brown}, {Castellani}, {Cocozza}, {De Luise}, {Delgado},
  {Ducourant}, {Galleti}, {Gilmore}, {Giuffrida}, {Holl}, {Kewley}, {Koposov},
  {Marinoni}, {Marrese}, {Osborne}, {Piersimoni}, {Portell}, {Pulone},
  {Ragaini}, {Sanna}, {Terrett}, {Walton}, {Wevers}, \&
  {Wyrzykowski}}]{Evans2018}
{Evans}, D.~W., {Riello}, M., {De Angeli}, F., {et~al.} 2018, \aap, 616, A4

\bibitem[{{Fabregat} \& {Reglero}(1990)}]{fabregat90}
{Fabregat}, J. \& {Reglero}, V. 1990, \mnras, 247, 407

\bibitem[{{Fender} {et~al.}(2004){Fender}, {Belloni}, \& {Gallo}}]{fender04}
{Fender}, R.~P., {Belloni}, T.~M., \& {Gallo}, E. 2004, \mnras, 355, 1105

\bibitem[{{Ferrigno} {et~al.}(2013){Ferrigno}, {Farinelli}, {Bozzo},
  {Pottschmidt}, {Klochkov}, \& {Kretschmar}}]{ferrigno13}
{Ferrigno}, C., {Farinelli}, R., {Bozzo}, E., {et~al.} 2013, \aap, 553, A103

\bibitem[{{Fitzgerald}(1970)}]{fitzgerald70}
{Fitzgerald}, M.~P. 1970, \aap, 4, 234

\bibitem[{{Friedman} {et~al.}(2011){Friedman}, {York}, {McCall}, {Dahlstrom},
  {Sonnentrucker}, {Welty}, {Drosback}, {Hobbs}, {Rachford}, \&
  {Snow}}]{friedman11}
{Friedman}, S.~D., {York}, D.~G., {McCall}, B.~J., {et~al.} 2011, \apj, 727, 33

\bibitem[{{Galazutdinov} {et~al.}(2000){Galazutdinov}, {Musaev},
  {Kre{\l}owski}, \& {Walker}}]{galazutdinov00}
{Galazutdinov}, G.~A., {Musaev}, F.~A., {Kre{\l}owski}, J., \& {Walker},
  G.~A.~H. 2000, \pasp, 112, 648

\bibitem[{{Gray} \& {Corbally}(2009)}]{gray09}
{Gray}, R.~O. \& {Corbally}, Christopher, J. 2009, {Stellar Spectral
  Classification}

\bibitem[{{Grundstrom} {et~al.}(2007){Grundstrom}, {Boyajian}, {Finch}, {Gies},
  {Huang}, {McSwain}, {O'Brien}, {Riddle}, {Trippe}, {Williams}, {Wingert}, \&
  {Zaballa}}]{grundstrom07b}
{Grundstrom}, E.~D., {Boyajian}, T.~S., {Finch}, C., {et~al.} 2007, \apj, 660,
  1398

\bibitem[{{Gutierrez-Moreno}(1979)}]{gutierrez79}
{Gutierrez-Moreno}, A. 1979, \pasp, 91, 299

\bibitem[{{Haigh} {et~al.}(2004){Haigh}, {Coe}, \& {Fabregat}}]{haigh04}
{Haigh}, N.~J., {Coe}, M.~J., \& {Fabregat}, J. 2004, \mnras, 350, 1457

\bibitem[{{Hanuschik}(1986)}]{hanuschik86}
{Hanuschik}, R.~W. 1986, \aap, 166, 185

\bibitem[{{Herbig}(1975)}]{herbig75}
{Herbig}, G.~H. 1975, \apj, 196, 129

\bibitem[{{Herbig} \& {Leka}(1991)}]{herbig91}
{Herbig}, G.~H. \& {Leka}, K.~D. 1991, \apj, 382, 193

\bibitem[{{Hummel} \& {Hanuschik}(1997)}]{hummel97}
{Hummel}, W. \& {Hanuschik}, R.~W. 1997, \aap, 320, 852

\bibitem[{{Hummel} \& {Vrancken}(1995)}]{hummel95}
{Hummel}, W. \& {Vrancken}, M. 1995, \aap, 302, 751

\bibitem[{{Humphreys} \& {McElroy}(1984)}]{humphreys84}
{Humphreys}, R.~M. \& {McElroy}, D.~B. 1984, \apj, 284, 565

\bibitem[{{Jaisawal} {et~al.}(2019){Jaisawal}, {Wilson-Hodge}, {Fabian},
  {Naik}, {Chakrabarty}, {Kretschmar}, {Ballantyne}, {Ludlam}, {Chenevez},
  {Altamirano}, {Arzoumanian}, {F{\"u}rst}, {Gendreau}, {Guillot}, {Malacaria},
  {Miller}, {Stevens}, \& {Wolff}}]{jaisawal19}
{Jaisawal}, G.~K., {Wilson-Hodge}, C.~A., {Fabian}, A.~C., {et~al.} 2019, \apj,
  885, 18

\bibitem[{{Janot Pacheco} {et~al.}(1982){Janot Pacheco}, {Chevalier}, \&
  {Ilovaisky}}]{janot82}
{Janot Pacheco}, E., {Chevalier}, C., \& {Ilovaisky}, S.~A. 1982, in IAU
  Symposium, Vol.~98, Be Stars, ed. M.~{Jaschek} \& H.-G. {Groth}, 151--154

\bibitem[{{Janot-Pacheco} {et~al.}(1981){Janot-Pacheco}, {Ilovaisky}, \&
  {Chevalier}}]{janot81}
{Janot-Pacheco}, E., {Ilovaisky}, S.~A., \& {Chevalier}, C. 1981, \aap, 99, 274

\bibitem[{{Jaschek} \& {G{\'o}mez}(1998)}]{jaschek98}
{Jaschek}, C. \& {G{\'o}mez}, A.~E. 1998, Highlights of Astronomy, 11, 566

\bibitem[{{Jayasinghe} {et~al.}(2019){Jayasinghe}, {Stanek}, {Kochanek},
  {Shappee}, {Holoien}, {Thompson}, {Prieto}, {Dong}, {Pawlak}, {Pejcha},
  {Shields}, {Pojmanski}, {Otero}, {Britt}, \& {Will}}]{jayasinghe19}
{Jayasinghe}, T., {Stanek}, K.~Z., {Kochanek}, C.~S., {et~al.} 2019, \mnras,
  486, 1907

\bibitem[{{Jenke} \& {Wilson-Hodge}(2017)}]{jenke17}
{Jenke}, P. \& {Wilson-Hodge}, C.~A. 2017, The Astronomer's Telegram, 10812, 1

\bibitem[{{Johnson}(1966)}]{johnson66}
{Johnson}, H.~L. 1966, \araa, 4, 193

\bibitem[{{Kennea} {et~al.}(2017){Kennea}, {Lien}, {Krimm}, {Cenko}, \&
  {Siegel}}]{kennea17}
{Kennea}, J.~A., {Lien}, A.~Y., {Krimm}, H.~A., {Cenko}, S.~B., \& {Siegel},
  M.~H. 2017, The Astronomer's Telegram, 10809, 1

\bibitem[{{Kos} \& {Zwitter}(2013)}]{kos13}
{Kos}, J. \& {Zwitter}, T. 2013, \apj, 774, 72

\bibitem[{{Kouroubatzakis} {et~al.}(2017){Kouroubatzakis}, {Reig}, {Andrews},
  \& {Zezas}}]{kouroubatzakis17}
{Kouroubatzakis}, K., {Reig}, P., {Andrews}, J., \& {Zezas}, A. 2017, The
  Astronomer's Telegram, 10822, 1

\bibitem[{{Landolt}(2009)}]{landolt09}
{Landolt}, A.~U. 2009, \aj, 137, 4186

\bibitem[{{Lenz} \& {Breger}(2005)}]{lenz05}
{Lenz}, P. \& {Breger}, M. 2005, Communications in Asteroseismology, 146, 53

\bibitem[{{Lyubimkov} {et~al.}(1997){Lyubimkov}, {Rostopchin}, {Roche}, \&
  {Tarasov}}]{lyubimkov97}
{Lyubimkov}, L.~S., {Rostopchin}, S.~I., {Roche}, P., \& {Tarasov}, A.~E. 1997,
  \mnras, 286, 549

\bibitem[{{Mainzer} {et~al.}(2011){Mainzer}, {Bauer}, {Grav}, {Cutri},
  {Dailey}, {Masiero}, {McMillan}, {Walker}, {Wright}, \& {Tholen}}]{mainzer11}
{Mainzer}, A.~K., {Bauer}, J.~M., {Grav}, T., {et~al.} 2011, in Lunar and
  Planetary Science Conference, Lunar and Planetary Science Conference, 1121

\bibitem[{{Martin} \& {Franchini}(2019)}]{martin19}
{Martin}, R.~G. \& {Franchini}, A. 2019, \mnras, 489, 1797

\bibitem[{{Martin} {et~al.}(2014){Martin}, {Nixon}, {Armitage}, {Lubow}, \&
  {Price}}]{martin14a}
{Martin}, R.~G., {Nixon}, C., {Armitage}, P.~J., {Lubow}, S.~H., \& {Price},
  D.~J. 2014, \apjl, 790, L34

\bibitem[{{Martin} {et~al.}(2011){Martin}, {Pringle}, {Tout}, \&
  {Lubow}}]{martin11}
{Martin}, R.~G., {Pringle}, J.~E., {Tout}, C.~A., \& {Lubow}, S.~H. 2011,
  \mnras, 416, 2827

\bibitem[{{Martins} {et~al.}(2005){Martins}, {Schaerer}, \&
  {Hillier}}]{martins05}
{Martins}, F., {Schaerer}, D., \& {Hillier}, D.~J. 2005, \aap, 436, 1049

\bibitem[{{Migliari} \& {Fender}(2006)}]{migliari06}
{Migliari}, S. \& {Fender}, R.~P. 2006, \mnras, 366, 79

\bibitem[{{Moritani} {et~al.}(2011){Moritani}, {Nogami}, {Okazaki}, {Imada},
  {Kambe}, {Honda}, {Hashimoto}, \& {Ichikawa}}]{moritani11}
{Moritani}, Y., {Nogami}, D., {Okazaki}, A.~T., {et~al.} 2011, \pasj, 63, 25

\bibitem[{{Moritani} {et~al.}(2013){Moritani}, {Nogami}, {Okazaki}, {Imada},
  {Kambe}, {Honda}, {Hashimoto}, {Mizoguchi}, {Kanda}, {Sadakane}, \&
  {Ichikawa}}]{moritani13}
{Moritani}, Y., {Nogami}, D., {Okazaki}, A.~T., {et~al.} 2013, \pasj, 65, 83

\bibitem[{{Negueruela} \& {Okazaki}(2001)}]{negueruela01a}
{Negueruela}, I. \& {Okazaki}, A.~T. 2001, \aap, 369, 108

\bibitem[{{Negueruela} {et~al.}(2001){Negueruela}, {Okazaki}, {Fabregat},
  {Coe}, {Munari}, \& {Tomov}}]{negueruela01b}
{Negueruela}, I., {Okazaki}, A.~T., {Fabregat}, J., {et~al.} 2001, \aap, 369,
  117

\bibitem[{{Negueruela} {et~al.}(1999){Negueruela}, {Roche}, {Fabregat}, \&
  {Coe}}]{negueruela99}
{Negueruela}, I., {Roche}, P., {Fabregat}, J., \& {Coe}, M.~J. 1999, \mnras,
  307, 695

\bibitem[{{Okazaki} {et~al.}(2013){Okazaki}, {Hayasaki}, \&
  {Moritani}}]{okazaki13}
{Okazaki}, A.~T., {Hayasaki}, K., \& {Moritani}, Y. 2013, \pasj, 65, 41

\bibitem[{{Parkes} {et~al.}(1980){Parkes}, {Murdin}, \& {Mason}}]{parkes80}
{Parkes}, G.~E., {Murdin}, P.~G., \& {Mason}, K.~O. 1980, \mnras, 190, 537

\bibitem[{{Pecaut} \& {Mamajek}(2013)}]{pecaut13}
{Pecaut}, M.~J. \& {Mamajek}, E.~E. 2013, \apjs, 208, 9

\bibitem[{{Porter}(1998)}]{porter98}
{Porter}, J.~M. 1998, \aap, 336, 966

\bibitem[{{Reig} {et~al.}(2017){Reig}, {Blay}, \& {Blinov}}]{reig17}
{Reig}, P., {Blay}, P., \& {Blinov}, D. 2017, \aap, 598, A16

\bibitem[{{Reig} \& {Blinov}(2018)}]{reig18b}
{Reig}, P. \& {Blinov}, D. 2018, \aap, 619, A19

\bibitem[{{Reig} \& {Fabregat}(2015)}]{reig15}
{Reig}, P. \& {Fabregat}, J. 2015, \aap, 574, A33

\bibitem[{{Reig} {et~al.}(1997){Reig}, {Fabregat}, {Coe}, {Roche},
  {Chakrabarty}, {Negueruela}, \& {Steele}}]{reig97b}
{Reig}, P., {Fabregat}, J., {Coe}, M.~J., {et~al.} 1997, \aap, 322, 183

\bibitem[{{Reig} {et~al.}(2007){Reig}, {Larionov}, {Negueruela}, {Arkharov}, \&
  {Kudryavtseva}}]{reig07b}
{Reig}, P., {Larionov}, V., {Negueruela}, I., {Arkharov}, A.~A., \&
  {Kudryavtseva}, N.~A. 2007, \aap, 462, 1081

\bibitem[{{Reig} {et~al.}(2001){Reig}, {Negueruela}, {Buckley}, {Coe},
  {Fabregat}, \& {Haigh}}]{reig01}
{Reig}, P., {Negueruela}, I., {Buckley}, D.~A.~H., {et~al.} 2001, \aap, 367,
  266

\bibitem[{{Reig} {et~al.}(2004){Reig}, {Negueruela}, {Fabregat}, {Chato},
  {Blay}, \& {Mavromatakis}}]{reig04}
{Reig}, P., {Negueruela}, I., {Fabregat}, J., {et~al.} 2004, \aap, 421, 673

\bibitem[{{Reig} {et~al.}(2005){Reig}, {Negueruela}, {Fabregat}, {Chato}, \&
  {Coe}}]{reig05b}
{Reig}, P., {Negueruela}, I., {Fabregat}, J., {Chato}, R., \& {Coe}, M.~J.
  2005, \aap, 440, 1079

\bibitem[{{Reig} {et~al.}(2016){Reig}, {Nersesian}, {Zezas}, {Gkouvelis}, \&
  {Coe}}]{reig16}
{Reig}, P., {Nersesian}, A., {Zezas}, A., {Gkouvelis}, L., \& {Coe}, M.~J.
  2016, \aap, 590, A122

\bibitem[{{Reig} \& {Zezas}(2014)}]{reig14a}
{Reig}, P. \& {Zezas}, A. 2014, \aap, 561, A137

\bibitem[{{Reig} {et~al.}(2010){Reig}, {Zezas}, \& {Gkouvelis}}]{reig10b}
{Reig}, P., {Zezas}, A., \& {Gkouvelis}, L. 2010, \aap, 522, A107

\bibitem[{{Riquelme} {et~al.}(2012){Riquelme}, {Torrej{\'o}n}, \&
  {Negueruela}}]{riquelme12}
{Riquelme}, M.~S., {Torrej{\'o}n}, J.~M., \& {Negueruela}, I. 2012, \aap, 539,
  A114

\bibitem[{{Rivinius} {et~al.}(2013){Rivinius}, {Carciofi}, \&
  {Martayan}}]{rivinius13a}
{Rivinius}, T., {Carciofi}, A.~C., \& {Martayan}, C. 2013, \aapr, 21, 69

\bibitem[{{Schlafly} \& {Finkbeiner}(2011)}]{schlafly11}
{Schlafly}, E.~F. \& {Finkbeiner}, D.~P. 2011, \apj, 737, 103

\bibitem[{{Schlegel} {et~al.}(1998){Schlegel}, {Finkbeiner}, \&
  {Davis}}]{schlegel98}
{Schlegel}, D.~J., {Finkbeiner}, D.~P., \& {Davis}, M. 1998, \apj, 500, 525

\bibitem[{{Shappee} {et~al.}(2014){Shappee}, {Prieto}, {Stanek}, {Kochanek},
  {Holoien}, {Jencson}, {Basu}, {Beacom}, {Szczygiel}, {Pojmanski},
  {Brimacombe}, {Dubberley}, {Elphick}, {Foale}, {Hawkins}, {Mullins},
  {Rosing}, {Ross}, \& {Walker}}]{shappee14}
{Shappee}, B., {Prieto}, J., {Stanek}, K.~Z., {et~al.} 2014, in American
  Astronomical Society Meeting Abstracts, Vol. 223, American Astronomical
  Society Meeting Abstracts \#223, 236.03

\bibitem[{{Silaj} {et~al.}(2010){Silaj}, {Jones}, {Tycner}, {Sigut}, \&
  {Smith}}]{silaj10}
{Silaj}, J., {Jones}, C.~E., {Tycner}, C., {Sigut}, T.~A.~A., \& {Smith}, A.~D.
  2010, \apjs, 187, 228

\bibitem[{{Slettebak}(1979)}]{slettebak79}
{Slettebak}, A. 1979, \ssr, 23, 541

\bibitem[{{Staubert} {et~al.}(2011){Staubert}, {Pottschmidt}, {Doroshenko},
  {Wilms}, {Suchy}, {Rothschild}, \& {Santangelo}}]{staubert11}
{Staubert}, R., {Pottschmidt}, K., {Doroshenko}, V., {et~al.} 2011, \aap, 527,
  A7

\bibitem[{{Steele} {et~al.}(1999){Steele}, {Negueruela}, \& {Clark}}]{steele99}
{Steele}, I.~A., {Negueruela}, I., \& {Clark}, J.~S. 1999, \aaps, 137, 147

\bibitem[{{Stevens} {et~al.}(1997){Stevens}, {Reig}, {Coe}, {Buckley},
  {Fabregat}, \& {Steele}}]{stevens97}
{Stevens}, J.~B., {Reig}, P., {Coe}, M.~J., {et~al.} 1997, \mnras, 288, 988

\bibitem[{{Sugizaki} {et~al.}(2020){Sugizaki}, {Oeda}, {Kawai}, {Mihara},
  {Makishima}, \& {Nakajima}}]{sugizaki20}
{Sugizaki}, M., {Oeda}, M., {Kawai}, N., {et~al.} 2020, arXiv e-prints,
  arXiv:2005.07971

\bibitem[{{Tao} {et~al.}(2019){Tao}, {Feng}, {Zhang}, {Bu}, {Zhang}, {Qu}, \&
  {Zhang}}]{tao19}
{Tao}, L., {Feng}, H., {Zhang}, S., {et~al.} 2019, \apj, 873, 19

\bibitem[{{Tsygankov} {et~al.}(2018){Tsygankov}, {Doroshenko}, {Mushtukov},
  {Lutovinov}, \& {Poutanen}}]{tsygankov18}
{Tsygankov}, S.~S., {Doroshenko}, V., {Mushtukov}, A. e.~A., {Lutovinov},
  A.~A., \& {Poutanen}, J. 2018, \mnras, 479, L134

\bibitem[{{Vacca} {et~al.}(1996){Vacca}, {Garmany}, \& {Shull}}]{vacca96}
{Vacca}, W.~D., {Garmany}, C.~D., \& {Shull}, J.~M. 1996, \apj, 460, 914

\bibitem[{{van den Eijnden} {et~al.}(2018){van den Eijnden}, {Degenaar},
  {Russell}, {Wijnand s}, {Miller-Jones}, {Sivakoff}, \& {Hern{\'a}ndez
  Santisteban}}]{eijnden18}
{van den Eijnden}, J., {Degenaar}, N., {Russell}, T.~D., {et~al.} 2018, \nat,
  562, 233

\bibitem[{{van den Eijnden} {et~al.}(2019){van den Eijnden}, {Degenaar},
  {Schulz}, {Nowak}, {Wijnands}, {Russell}, {Hern{\'a}ndez Santisteban},
  {Bahramian}, {Maccarone}, {Kennea}, \& {Heinke}}]{eijnden19}
{van den Eijnden}, J., {Degenaar}, N., {Schulz}, N.~S., {et~al.} 2019, \mnras,
  487, 4355

\bibitem[{{Walborn} \& {Fitzpatrick}(1990)}]{walborn90}
{Walborn}, N.~R. \& {Fitzpatrick}, E.~L. 1990, \pasp, 102, 379

\bibitem[{{Wegner}(1994)}]{wegner94}
{Wegner}, W. 1994, \mnras, 270, 229

\bibitem[{{Wegner}(2006)}]{wegner06}
{Wegner}, W. 2006, \mnras, 371, 185

\bibitem[{{Wilson-Hodge} {et~al.}(2018){Wilson-Hodge}, {Malacaria}, {Jenke},
  {Jaisawal}, {Kerr}, {Wolff}, {Arzoumanian}, {Chakrabarty}, {Doty},
  {Gendreau}, {Guillot}, {Ho}, {LaMarr}, {Markwardt}, {{\"O}zel}, {Prigozhin},
  {Ray}, {Ramos-Lerate}, {Remillard}, {Strohmayer}, {Vezie}, {Wood}, \& {NICER
  Science Team}}]{wilson18}
{Wilson-Hodge}, C.~A., {Malacaria}, C., {Jenke}, P.~A., {et~al.} 2018, \apj,
  863, 9

\bibitem[{{Wright} {et~al.}(2010){Wright}, {Eisenhardt}, {Mainzer}, {Ressler},
  {Cutri}, {Jarrett}, {Kirkpatrick}, {Padgett}, {McMillan}, {Skrutskie},
  {Stanford}, {Cohen}, {Walker}, {Mather}, {Leisawitz}, {Gautier}, {McLean},
  {Benford}, {Lonsdale}, {Blain}, {Mendez}, {Irace}, {Duval}, {Liu}, {Royer},
  {Heinrichsen}, {Howard}, {Shannon}, {Kendall}, {Walsh}, {Larsen}, {Cardon},
  {Schick}, {Schwalm}, {Abid}, {Fabinsky}, {Naes}, \& {Tsai}}]{wright10}
{Wright}, E.~L., {Eisenhardt}, P. R.~M., {Mainzer}, A.~K., {et~al.} 2010, \aj,
  140, 1868

\bibitem[{{Zhang} {et~al.}(2005){Zhang}, {Qu}, {Song}, \& {Torres}}]{zhang05}
{Zhang}, S., {Qu}, J.-L., {Song}, L.-M., \& {Torres}, D.~F. 2005, \apjl, 630,
  L65

\bibitem[{{Zhang} {et~al.}(2019){Zhang}, {Ge}, {Song}, {Zhang}, {Qu}, {Zhang},
  {Doroshenko}, {Tao}, {Ji}, {G{\"u}ng{\"o}r}, {Santangelo}, {Shi}, {Chang},
  {Chen}, {Chen}, {Chen}, {Chen}, {Chen}, {Cui}, {Cui}, {Deng}, {Dong}, {Du},
  {Fu}, {Gao}, {Gao}, {Gao}, {Gu}, {Guan}, {Guo}, {Han}, {Hu}, {Huang}, {Huo},
  {Jia}, {Jiang}, {Jiang}, {Jin}, {Jin}, {Li}, {Li}, {Li}, {Li}, {Li}, {Li},
  {Li}, {Li}, {Li}, {Li}, {Li}, {Liang}, {Liao}, {Liu}, {Liu}, {Liu}, {Liu},
  {Liu}, {Liu}, {Liu}, {Lu}, {Lu}, {Luo}, {Ma}, {Meng}, {Nang}, {Nie}, {Ou},
  {Sai}, {Sun}, {Tan}, {Tao}, {Tuo}, {Wang}, {Wang}, {Wang}, {Wang}, {Wang},
  {Wen}, {Wu}, {Wu}, {Xiao}, {Xiong}, {Xu}, {Xu}, {Yan}, {Yang}, {Yang},
  {Yang}, {Zhang}, {Zhang}, {Zhang}, {Zhang}, {Zhang}, {Zhang}, {Zhang},
  {Zhang}, {Zhang}, {Zhang}, {Zhang}, {Zhang}, {Zhang}, {Zhang}, {Zhang},
  {Zhao}, {Zhao}, {Zhao}, {Zheng}, {Zhu}, {Zhu}, {Zou}, \& {Insight-HXMT
  Collaboration}}]{zhang19}
{Zhang}, Y., {Ge}, M., {Song}, L., {et~al.} 2019, \apj, 879, 61

\end{thebibliography}

\end{document}